\documentclass{article}
\usepackage[utf8]{inputenc}
\usepackage{arabtex}
\usepackage{textcomp}
\usepackage[numbers]{natbib}
\usepackage{mathtools}
\usepackage{float}
\title{Characterization of the terahertz photoconductive antenna by three-dimensional finite-difference time-domain method}
\author{Jitao Zhang $^{\dag}$\\ ECE Department,The University of Arizona, Tucson, AZ,85721\\
$^{\dag}$ \textit{jitaozhang@email.arizona.edu}}

\date{%
    \today
    \\[2\baselineskip]
    \normalfont\normalsize%
    \parbox{0.8\linewidth}{%
{\bfseries Abstract}: We numerically describe the physical mechanism underlying the terahertz photoconductive antenna (PCA) by the finite-difference time-domain method in three-dimension. The feature of our approach is that the multi-physical phenomena happening in the PCA, such as light-matter interaction, photo-excited carrier dynamics and full-wave propagation of the THz radiation, are considered and embodied in the simulation. The method has been verified by comparing with existing commercial softwares. In addition, we use this simulation tool to characterize the parameter-dependent performance of a PCA,thereby the design of novel PCA with enhanced optics-to-THz efficiency can be inspired.
    }
}
\usepackage{natbib}
\usepackage{graphicx}

\begin{document}

\maketitle

\section{Introduction}

Terahertz (THz) photoconductive antenna (PCA) is one of the most commonly used devices as THz source and/or detector. It generates and detects THz radiation by transient photocarriers induced with ultrafast laser pulses\citep{austin1994}. Several routes have been explored to understand how a PCA works in the past several decades\citep{fdtdel1990,fdtdsano1991,fdtdsirbu2005,fdtdkira2009,fdtdnazeri2010,drudejepsen1996,drudepiao2000,drudeduv2001,ecmholzman2000,ecmezdi2006,ecmhy2013},based on different physical models. These can be cataloged into three approaches. 

The first approach is mainly simulating the photo-excited carrier dynamics inside the semiconductor using Drude-Lorentz model\citep{drudejepsen1996,drudepiao2000,drudeduv2001}. The Drude-Lorentz model is a very straightforward and simplified way to model the dynamics of the carrier transport inside the active semiconductor layer of a PCA. It is quite effective to analyze the dependence of the THz radiation on the material’s properties, such as carrier lifetime, mobility, doping density and absorption, as well as on the laser source’s intensity and pulse width. However, this method can only calculate the photo-excited current inside the semiconductor (in the near-field), but the far-field radiation can only be deduced in an approximate way. In other words, the different radiation properties of the antennas with various shapes cannot be distinguished by the Drude-Lorentz model. Moreover, this model can hardly simulate the space-related phenomena, such as the effect of asymmetrical illumination of the laser spot within the PCA’s gap\citep{sano1989}. 

The second approach is depicting the PCA as a special lumped element based on the equivalent circuit model (ECM)\citep{ecmholzman2000,ecmezdi2006,ecmhy2013}. In this case, a PCA can be considered as a combination of voltage or current source with time-varying resistance and antenna impedance. The laser-induced resistance of the source is evaluated by the carrier dynamics, and the power of THz radiation is deduced by means of the general antenna theory. This approach adopts the existing antenna theory for PCA's analysis, so that all of the antenna-related aspects of a PCA can be simulated and understood in deep. For example, based on the ECM, one can study the impedance-matching efficiency between the photo-exicted source and the radiation antenna, and even obtain the radiation properties of a PCA in far-field with the aid of strong commercial software (for example, HFSS).The above features of the ECM make it very suitable to study the antenna-related properties of a PCA and further help to design a PCA having better performance. Unfortunately, the assumptions made in the ECM limit its application. For instance, the gap of the electrodes is usually assumed to be fully and uniformly illuminated by the laser beam when calculating the source resistance. Therefore,the ECM cannot simulate the space-related phenomena in the near field either. Sometimes the exponential decay of the photocurrent caused by the recombination of photo-excited carriers is also ignored, hereby the laser pulse's information is lost as a consequence.

The third approach (full-wave model) couples the carriers dynamics with the full-wave interaction and propagation, which is believed to be superior to the other two on its equal capability to simulate the phenomena in both near field and far field\citep{fdtdel1990,fdtdsano1991,fdtdsirbu2005,fdtdkira2009,fdtdnazeri2010}. This approach can fulfill the needs of the comprehensive simulation, in which almost all of the parameters that tightly related to the performance of the PCA can be involved. 

In this work, we developed a computational algorithm to carry out the full-wave model by in-house codes based on the finite-difference time-domain (FDTD) method in three-dimension (3D). Further, we made an attempt to implement this simulation tool to characterize the parameter-dependent performance of a PCA, thereby the design of novel PCA with enhanced optics-to-THz efficiency can be inspired. The rest of the paper is arranged as follows. Section 2 explains the physical model utilized in this method. Section 3 describe the details about the numerical simulation methods. The results of the simulation are presented in Sec.4, and a conclusion is drawn in the end.

\section{Physical model}
A typical PCA consists of a semiconductor material and a pair of electrodes with a gap between them deposited on the semiconductor's surface, as shown in Fig.1. To generate THz radiation, the gap region of the semiconductor is illuminated by an ultrafast laser pulse (usually in hundred femtoseconds or less) and the electrodes are biased by a DC voltage. The photo-excited carriers (i.e. electrons and holes) inside the semiconductor will driven by the biased field to generate transient current, which then will be radiated into free space with the help of the electrodes that acting as an antenna.

\begin{figure}[h!]
\centering
\includegraphics[width=.8\textwidth]{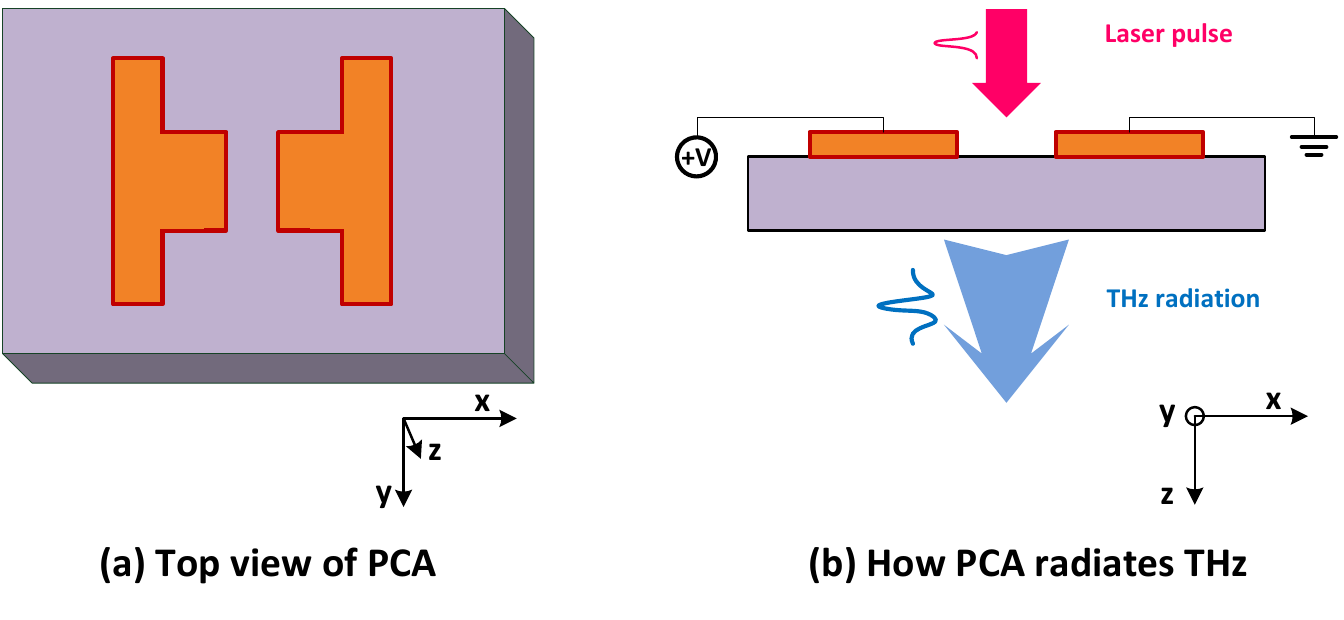}
\caption{Schematic structure of the PCA and THz radiation from the PCA. (a) A pair of dipole electrodes deposited on the surface of the semiconductor material, and (b) THz radiation from the biased PCA caused by the illumination of the laser pulse.}
\label{fig1:thz-tdssystem}
\end{figure}

The THz radiation process of a PCA can be understood in three phases, saying the build of static electric field, the generation of the photo-excited transient current in the near-field, and the THz radiation in both the near- and far-field.

In the first phase, a static electric field is built inside the bulk semiconductor when a DC voltage is biased on the electrodes. This static field will provide an initial field to drive the photo-excited carriers to flow towards the electrodes in the second phase. The Poisson equation associated with the carrier dynamics equations  (i.e. drift-diffusion equation and continuity equation) can be applied to solve this problem, as shown in Eq.(1) $\sim$ (5):

\begin{equation}
\nabla^{2}V(\vec{r})=\frac{q}{\varepsilon}(n(\vec{r})-p(\vec{r})-N_{D}+N_{A})
\end{equation}

\begin{equation}
\nabla\cdot J_{n}(\vec{r})=qR
\end{equation}

\begin{equation}
\nabla\cdot J_{p}(\vec{r})=-qR
\end{equation}

\begin{equation}
J_{n}(\vec{r})=q\mu_{n}n(\vec{r})(-\nabla\cdotp V(\vec{r}))+qD_{n}\nabla n(\vec{r})
\end{equation}

\begin{equation}
J_{p}(\vec{r})=q\mu_{p}p(\vec{r})(-\nabla\cdotp V(\vec{r}))-qD_{p}\nabla p(\vec{r}),
\end{equation}

\noindent where \textit{V} is the voltage distribution inside the semiconductor, \textit{q} is elementary charge, $\varepsilon $ is permittivity of the semiconductor, \textit{n} and \textit{p} are density of electrons and holes, respectively, $ N_{D}-N_{A} $ represents the concentration of impurities, $J_{n}$ and $J_{p}$ are current density of electrons and holes, respectively, $R$ is the recombination rate of the carriers, $\mu_{n}$ and $\mu_{p}$ are mobilities of the carriers, and $D_{n}$ and $D_{p}$ are diffusion coefficients, which are related to the mobilities by Einstein relationship
\begin{equation}
\frac{D_{n}}{\mu_{n}}=\frac{D_{p}}{\mu_{p}}=\frac{K_{B}T}{q}.
\end{equation}
The $\vec{r}$ indicates that the corresponding parameters are vectors. By solving the above equations, we can obtain the steady solution of the electric field ($E_{DC}$), carrier densities  ($n_{DC}$ and $p_{DC}$)  and current density ($J_{n_{DC}}$ and $J_{p_{DC}}$) inside the semiconductor for the first phase.

In the second phase, a transient current will be generated when a laser pulse illuminate the PCA's gap according to the carrier dynamics model. Then in the third phase, the transient current will result in THz radiation through the electrodes, which can be predicted by Maxwell's equation. The coupling between phase 2 and phase 3 is realized by using the transient current as driving source of the antenna to update the electromagnetic field. The physical model used here can be summarized as eq.(7) $\sim$ (14):

\begin{equation}
\nabla\times E(\vec{r})=-\mu \frac{\partial H(\vec{r}) }{\partial t}
\end{equation}

\begin{equation}
\nabla\times H(\vec{r})=\varepsilon \frac{\partial E(\vec{r}) }{\partial t}+J_{n,pho}(\vec{r})+J_{p,pho}(\vec{r})
\end{equation}

\begin{equation}
q\frac{\partial n(\vec{r})}{\partial t} = \nabla\cdot J_{n}(\vec{r})+q(G-R)
\end{equation}

\begin{equation}
q\frac{\partial p(\vec{r})}{\partial t} = - \nabla\cdot J_{p}(\vec{r})+q(G-R)
\end{equation}

\begin{equation}
J_{n}(\vec{r})=q\mu_{n}n(E_{DC}(\vec{r})+E(\vec{r}))+qD_{n}\nabla n(\vec{r})
\end{equation}

\begin{equation}
J_{p}(\vec{r})=q\mu_{p}p(E_{DC}(\vec{r})+E(\vec{r}))-qD_{p}\nabla p(\vec{r})
\end{equation}

\begin{equation}
J_{n,pho}(\vec{r})=J_{n}(\vec{r})-J_{n_{DC}}(\vec{r})
\end{equation}

\begin{equation}
J_{p,pho}(\vec{r})=J_{p}(\vec{r})-J_{p_{DC}}(\vec{r}),
\end{equation}

\noindent where $E$ and $H$ are radiated electric and magnetic field, respectively,$\mu$ is permeability, $J_{n,pho}$ and $J_{p,pho}$ specifically represent the photo-excited current density, and $G$ is the generation rate of the photo-excited carriers. Other symbols have the same meanings as above. However, it should be noted that the photo-excited effects are involved here. The continuity equation shown in Eq.(9) and (10) depicts the carrier dynamics, and drift-diffusion equation shown in Eq.(11) and (12) describes the corresponding transient current. By using photo-excited current (Eq.(13) and (14)) as driving source of an antenna to update the Maxwell's equation (Eq.(7) and (8)), the THz radiation can be precisely predicated, both in the near-field and far-field.

\section{Numerical simulation method}

\subsection{Steady-state solution of DC field by finite-difference method}
According to Sec.2, the built DC field when the electrodes are biased should be solved once before implementing the time-domain solution. The finite-difference method is applied to solve Poisson equation in 3D \cite{dcfich1983}. To carry out this, the carrier density is first described by the Boltzmann approximation in Eq.(15)-(16)

\begin{equation}
n=n_{i}\exp \left(\frac{q(V-V_{n})}{kT}\right)
\end{equation}

\begin{equation}
p=n_{i}\exp \left(\frac{q(V_{p}-V)}{kT}\right).
\end{equation}
Where $n_{i}$ is the intrinsic carrier density, $V$ is the biased voltage inside the material, $V_{n}$ and $V_{p}$ are \emph{quasi-Femi potentials} of $n$ and $p$, respectively. Under equilibrium condition (no bias), the quasi-Fermi levels are the same for electrons and holes so that $V_{n}=V_{p}$. It then can be immediately deduced from Eq. (15) and (16) that $np=n_{i}^2$ . When the semiconductor is biased, $V_{n} \neq V_{p}$ because the electron and hole concentrations will depart from their equilibrium value under nonequilibrium conditions. The recombination rate $R$ in Eq.(2)\&(3) (also for Eq.(7) \& (8)) can be described by the Shockley-Read-Hall (SRH) process 

\begin{equation}
R=\frac{np-n_{i}^2}{\tau_{p}n+\tau_{n}p},
\end{equation}
\noindent where $\tau_{n}$ and $\tau_{p}$ are carrier lifetimes.

Choosing $V$,$V_{n}$ and $V_{p}$ as variables, Eq.(1)$\sim$ (5) can be rewritten as

\begin{equation}
\nabla^{2}V=\frac{q}{\varepsilon} \left(n_{i}\exp \left(\frac{(V-V_{n})}{V_{T}}\right)-n_{i}\exp \left(\frac{(V_{p}-V)}{V_{T}}\right)-N_{D}+N_{A} \right)
\end{equation}

\begin{equation}
\nabla\cdot \left( \mu_{n}\exp \left(\frac{(V-V_{n})}{V_{T}}\right) (-\nabla\cdotp V)+D_{n}\nabla \left( \exp \left(\frac{(V-V_{n})}{V_{T}}\right) \right) \right) = \frac{R}{ n_{i}}
\end{equation}

\begin{equation}
\nabla\cdot \left( \mu_{p}\exp \left(\frac{(V_{p}-V)}{V_{T}}\right) (-\nabla\cdotp V)+D_{p}\nabla \left( \exp \left(\frac{(V_{p}-V)}{V_{T}}\right) \right) \right) = -\frac{R}{ n_{i}},
\end{equation}

\noindent where $ V_{T}= kT/q $ is the thermal voltage of the semiconductor. Each of these three equations is responsible for solving one variable by approximating the differential operation with finite-difference . For example, $V$ is solved first according to Eq.(18), and then $V_{n}$ and $V_{p}$ are solved according to Eq.(19) \& (20). The Gummel's algorithm is used to obtain the steady-state solution by iteration.

The interface between the electrodes and the semiconductor are considered as Ohmic contact, and Dirichlet boundary condition is applied. Other artificial boundaries of the semiconductor are considered as Neumann boundary. For the Dirichlet boundary condition at the electrode-semiconductor interface, thermal equilibrium and electric neutrality are assumed. It can be summarized as

\begin{equation}
n_{B}=\frac{k^{2}}{2}+\sqrt{\frac{k^{2}}{4}+n_{i}^{2}}
\end{equation}

\begin{equation}
p_{B}=n_{i}^{2}/n_{B}
\end{equation}
\begin{equation}
V_{B}=V_{bias},
\end{equation}
\noindent where $k=N_{D}-N_{A}$ denotes the concentration of impurities, $V_{bias}$ is the biased voltage, and the subscript $B$ indicates the value at the boundary.

When the solution at equilibrium condition is obtained, the voltage can be added step by step along with the iteration loop until the biased voltage is achieved. The step size of the voltage $\Delta V$ usually equals $V_{T}$ to avoid the instability of the algorithm. Once the calculation of DC field is completed, we will obtain the steady-state solutions of the electric field, carrier densities and current densities, which will be further used as input data of time-domain solution.

\subsection{Time-domain solution in near-field by FDTD}

When the steady-state solution for biased voltage is obtain, FDTD method is applied to the time-domain solution of the laser-induced THz radiation \cite{fdtdbook}. FDTD is one of the most powerful tools that can solve Maxwell's equation in time domain. The basic idea of the FDTD is to update electric field $E$ and magnetic field $H$ by \emph{leapfrog} manner within a meshed space (i.e. Yee's cell). It is realized by means of replacing the time-derivative operation in Maxwell's equation by finite-difference approximation with second-order accuracy. In our simulation, the electromagnetic fields are updated by a \emph{two-step} process \cite{fdtdbook}, and $J$,$n$ and $p$ are also updated accordingly. Basically, $E$ is updated at integral time step (for example, $ E^{t} $), and $H$ is updated at half-time step ($ H^{t+1/2} $). In addition, $J_{n}$ and $J_{p}$ should be updated at same time step as $E$ according to Eq.(11) \& (12) ($J_{n}^{t}$,$J_{p}^{t}$), and $n$ and $p$ should be updated at half-time step according to Eq. (9) \& (10) ($n^{t+1/2}$, $p^{t+1/2}$ ). It should be note that since the variables are updated at different time step, time-averaging approximation is usually used to implement this method. For example, the value of $J_{n}$ and $J_{p}$ at half-time step in Eq.(8) can be approximated as

\begin{equation}
J_{n}^{t+1/2}=\frac{J_{n}^{t}+J_{n}^{t+1}}{2}
\end{equation}

\begin{equation}
J_{p}^{t+1/2}=\frac{J_{p}^{t}+J_{p}^{t+1}}{2}.
\end{equation}
\noindent Similar approximation should be carried out for $n$ and $p$ in Eq.(11) \& (12)

\begin{equation}
n^{t}=\frac{n^{t-1/2}+n^{t+1/2}}{2}
\end{equation}

\begin{equation}
p^{t}=\frac{p^{t-1/2}+p^{t+1/2}}{2}.
\end{equation}
\noindent The completed discretization forms can be found in Appendix A.

The laser pulse is coupled into the simulation model by calculating the generation rate $G$ of the photo-excited carriers. Assuming that the laser beam has Gaussian shape both in temporal and spatial domain, $G$ is given as

\begin{align}
 G(\vec{r},t) = & \notag \frac{I_{0}}{h\nu} \cdot \exp \left(-\frac{(x-x_{0})^{2}}{\sigma_{x}^{2}} \right) \cdot \exp \left(-\frac{(y-y_{0})^{2}}{\sigma_{y}^{2}} \right) \\
 &  \cdot \exp \left(-\frac{4ln(2)(t-t_{0})^{2}}{\sigma_{t}^{2}} \right) \\
 & \notag \cdot \alpha \exp \left(-\alpha (z-z_{0}) \right),
\end{align}
\noindent where $I_{0}$ is the laser power intensity, $\alpha$ is the absorption coefficient of the semiconductor, $h$ is Planck constant, $\nu$ is optical frequency, $(x_{0},y_{0},z_{0})$ is the initial location of the laser beam (here the laser beam propagates in $z$-direction), $\sigma_{x}$ and $\sigma_{y}$ represent the beam waist, $t_{0}$ represents the temporal peak of the laser pulse, and $\sigma_{t}$ is the temporal full-width-half-maximum (FWHM) of the pulse (pulse duration).

In FDTD method, to simulate the propagation of the electromagnetic field in free-space by the limited computational memory, an absorption layer enclosing the computational region should be employed to prevent any reflection at the boundary. For this purpose, the \emph{uniaxial perfect-matched-layer} (UPML) is used here, which can be readily coupled into the codes.

 \subsection{Near-to-far-field transformation}
 Even though the FDTD method is a powerful solver of Maxwell's equation, the requirement of the huge computational memory, which increases dramatically with the computational region, usually limits its application to the near-field. Fortunately, the far-field radiation can be accurately deduced by the near-field result based on the equivalence principle \cite{ntff1991}. The basic idea is as follows. If one encloses the actual source by a closed surface $s$, the equivalent surface currents derived from the source's radiation can substitute the actual source to predict the far-field radiation. These currents are obtained by means of 
 
 \begin{equation}
 J_{s}(\vec{r},t)= \hat{n}\times H(\vec{r},t)
 \end{equation}
 
 \begin{equation}
 M_{s}(\vec{r},t)= -\hat{n}\times E(\vec{r},t),
 \end{equation} 
\noindent where $J_{s}$ is magnetic surface current density, $M_{s}$ is electric surface current density, $\hat{n}$ is a unit vector normal to the surface $S$ and coming out of it, and $E$ and $H$ are near-field radiation. The far-field radiation can be predicated by the equivalent surface current in time-domain by means of 

\begin{equation}
  E_{r}(\vec{r},t)\cong 0
\end{equation}

\begin{equation}
  E_{\theta}(\vec{r},t)=-\eta_{0} W_{\theta}(\vec{r},t)-U_{\phi}(\vec{r},t)
\end{equation}
 
\begin{equation}
  E_{\phi}(\vec{r},t)=-\eta_{0} W_{\phi}(\vec{r},t)+U_{\theta}(\vec{r},t),
\end{equation}

\noindent where $\eta_{0} = \sqrt{\mu_{0}/\varepsilon_{0}}$ is the impedance of free space, and $W$ and $U$ are yielded by

\begin{equation}
  W(\vec{r},t)= \frac{1}{4\pi r c}\frac{\partial}{\partial t}\left[ \oint J_{s} \left(t-\frac{r-r'\cdot \hat{r}}{c}\right)\cdot ds'   \right]
\end{equation}
 
\begin{equation}
 U(\vec{r},t)= \frac{1}{4\pi r c}\frac{\partial}{\partial t}\left[ \oint M_{s} \left(t-\frac{r-r'\cdot \hat{r}}{c}\right)\cdot ds'   \right],
\end{equation} 
\noindent where $r$ is the distance of the far-field point to the origin, $r'$ is the distance of the near-field point to the origin, $\hat{r}$ is the unit direction vector. The integral is performed on the whole equivalent surface $s$. According to the Eq.(31)-(35), the calculation of the radiation in the far-field point can be embedded in the FDTD's time-loop so that it can be updated simultaneously. However, since each point on the equivalent surface has different time delay to far-field point, their contributions to the far-field point will be delayed according to their positions for each time loop. This problem can be solved by designate a void time series with sufficient length for the far-field point in advance \cite{ntff1991}.

\section{Results and discussions}
\subsection{Verification of the method}
The validity of the simulation method is verified by comparing with commercial softwares. To date, there does not exist any commercial software that can model the comprehensive interaction listed in the physical modeling of a PCA all-in-one. Fortunately, due to the flexibility of the codings, we can separate the codes of the simulation method into three parts based on Sec.3 and verify them separately. For example, Sec.3.1 and 3.2 are verified by comparing with SILVACO and COMSOL, and Sec.3.3 is verified by comparing with HFSS. In all of the comparisons mentioned above,the identical input data are used and the results are compared directly.

\begin{figure}[H]
\centering
\includegraphics[width=1\textwidth]{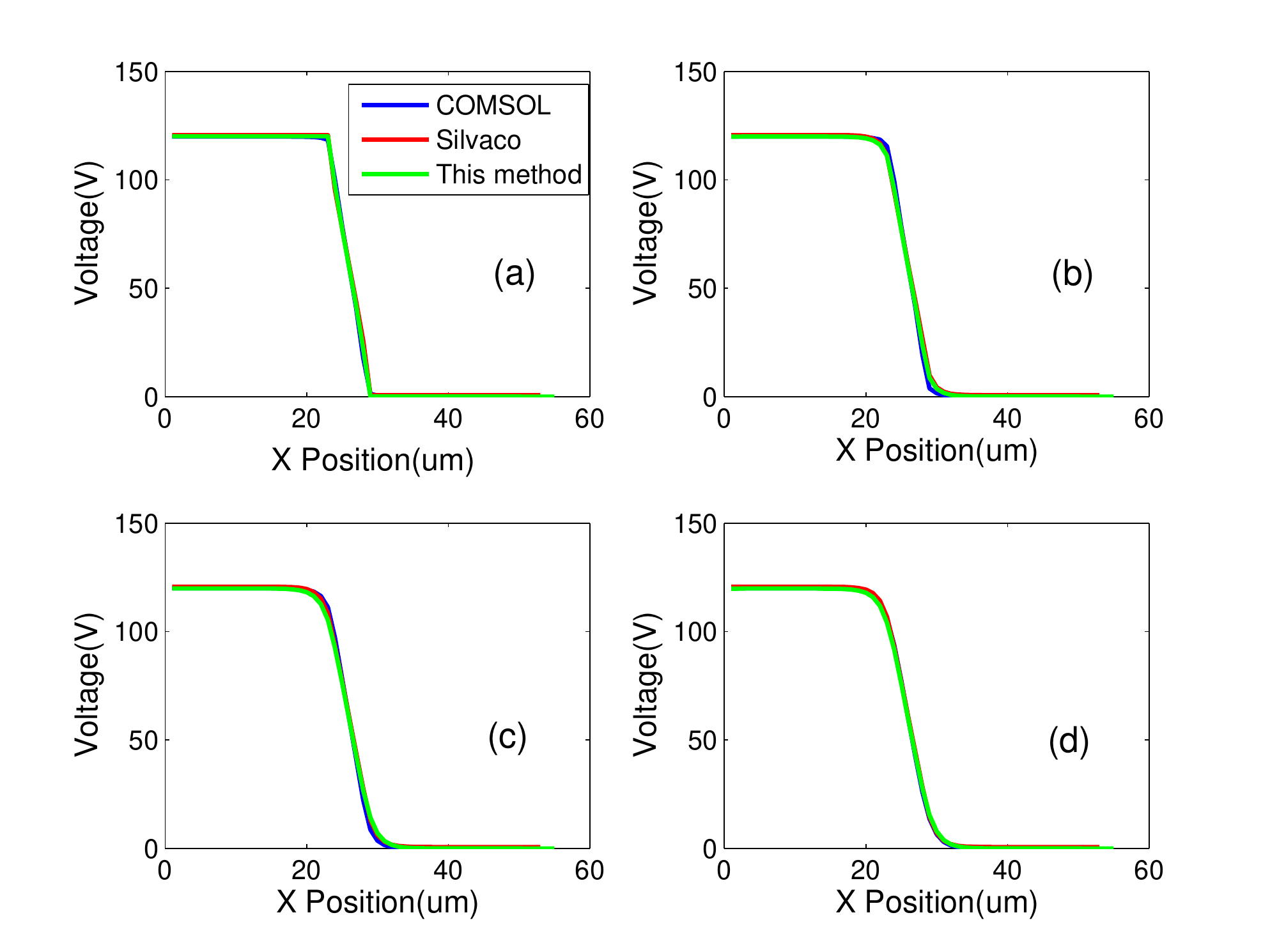}
\caption{ DC voltage comparison between this method, SILVACO and COMSOL. The structure of the PCA is the same as Figure \ref{fig1:thz-tdssystem}. The bias voltage is set as $120V$, and the solutions are compared along the line that passing the center of the PCA and parallel to x-axis in $XOY$ plane at depths of (a)$0um$, (b)$0.8um$, (c)$1.6um$ and (d)$2um$.}
\label{fig:dc_MSC}
\end{figure}

For the comparison of the steady-state solution of DC field, the same dipole PCA is simulated by in this method, SILVACO and COMSOL. The simulation is accomplished by TCAD tool and Semiconductor Module in SILVACO and COMSOL, respectively. The DC voltage across the gap of the PCA are compared at different depths in Figure \ref{fig:dc_MSC}. The maximum discrepancy between them is less than $6\%$, which indicates the accuracy of this method.

\begin{figure}[H]
\centering
\includegraphics[width=1\textwidth]{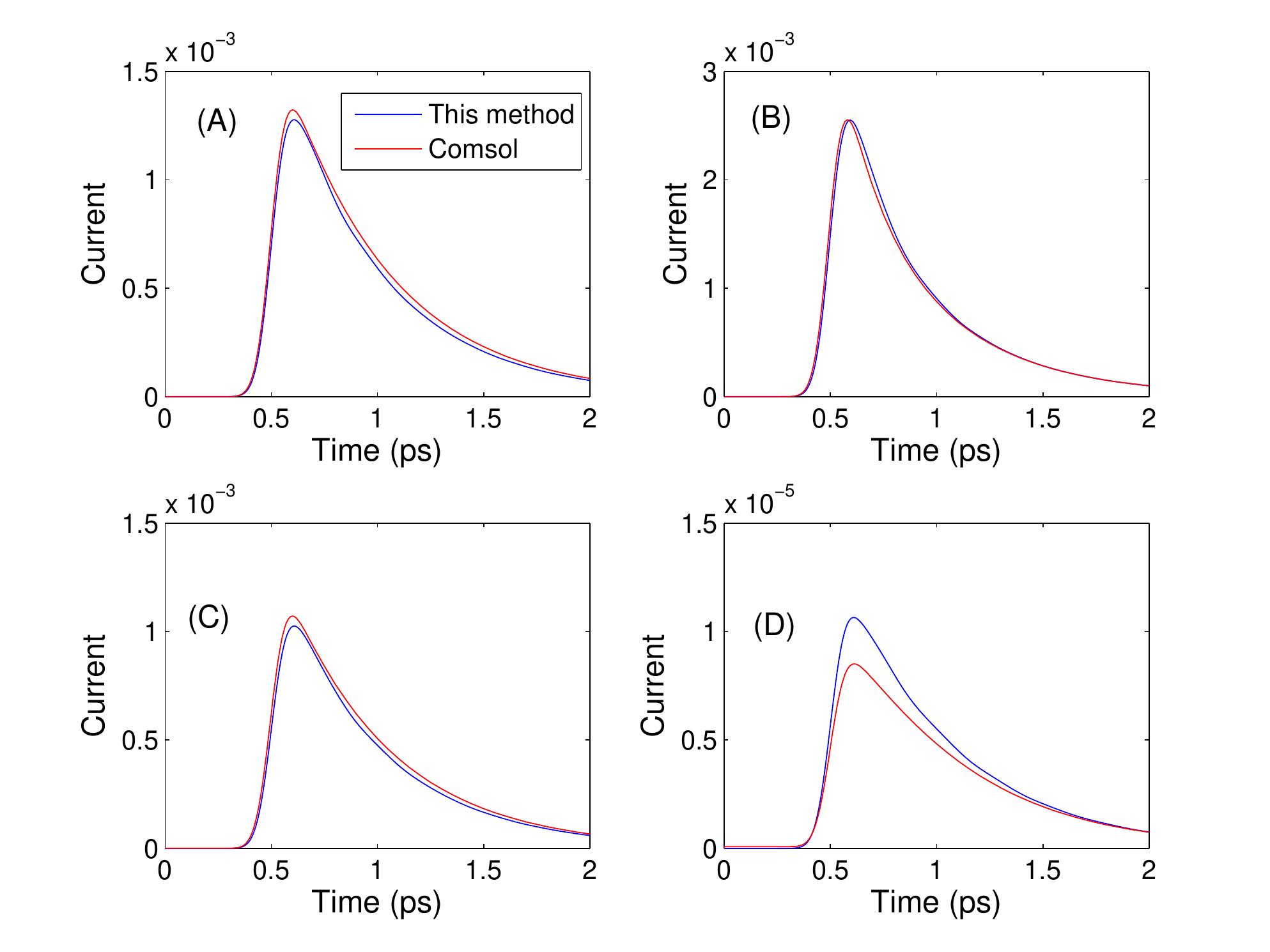}
\caption{Comparison of Photo-excited currents of the cross-sections at the locations of (A)the vicinity of the anode,(B)the center of the laser beam, (C)the right edge of the beam and (D) the vicinity of cathode.}
\label{fig:nf_MC}
\end{figure}

In addition, for the comparison of time-domain solution in near-field, the photo-excited current within the gap of the PCA are calculated, by both this method and COMSOL. To consider a general case, a coplanar strip-line type PCA are chosen. Two identical rectangular electrodes with length of $50um$ and width of $5um$ are placed on LT-GaAs substrate and parallel to each other, with a gap of $34um$. The gap is partially illuminated by a $20um$ Gaussian laser beam located at the vicinity of the anode. The pulse width of the laser is $80fs$, and the average power is $2.6mw$. The bias voltage is $5V$. The transient photo-excited currents of the cross-sections perpendicular to the bias field at different locations of the gap are calculated and compared. Figure \ref{fig:nf_MC} shows the result. It indicates the consistency between this method and COMSOL, and the discrepancy is mainly caused by the mesh error.

\begin{figure}[H]
\centering
\includegraphics[width=1\textwidth]{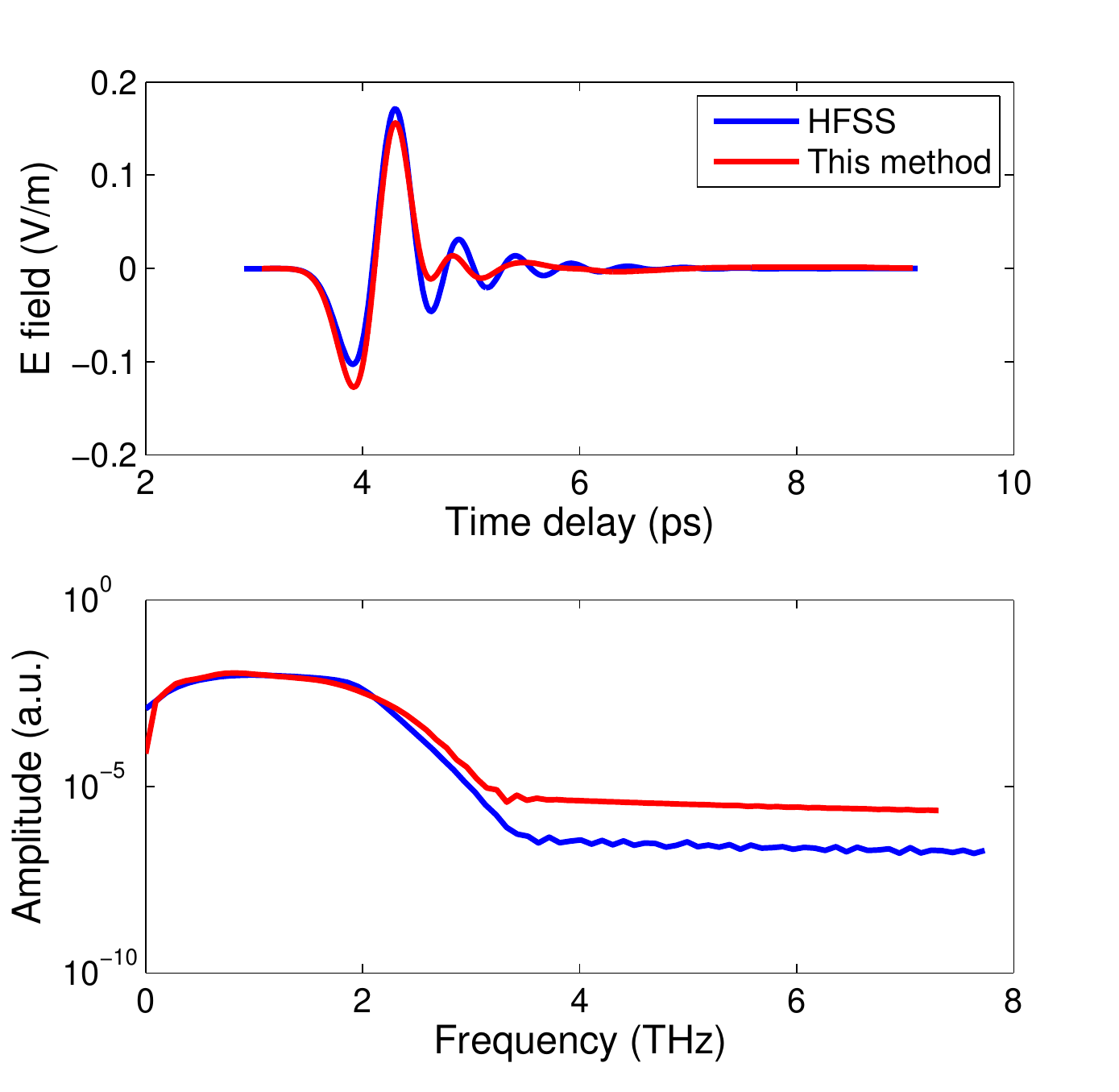}
\caption{Comparison of the near-to-far-field transformation.(upper) Time-domain far-field radiation,(lower) Corresponding spectrum by means of Fourier transform.}
\label{fig:far_MH}
\end{figure}

In addition, the near-to-far-field transformation of this method is verified by HFSS. A Gaussian-shaped current is applied as hard source of the PCA structure for both this method and HFSS, and the far-field radiation is compared directly. The results are shown in Figure \ref{fig:far_MH}. It indicates that main pulse in time-domain calculated by both methods are very close, and there is a few of discrepancy after the main pulse, which is probably caused by the numerical error in this method (e.g. the mesh size is not small enough). In addition, the frequency responses also agrees with each other, except the discrepancy beyond $3THz$.

In short, the comparison results shown in Figure 2,3\&4 confirm the validity of this method. Further verification will be made by comparing the simulation result with the experiment data.

\subsection{Simulation results of a dipole PCA}
\begin{figure}[H]
\centering
\includegraphics[width=.8\textwidth]{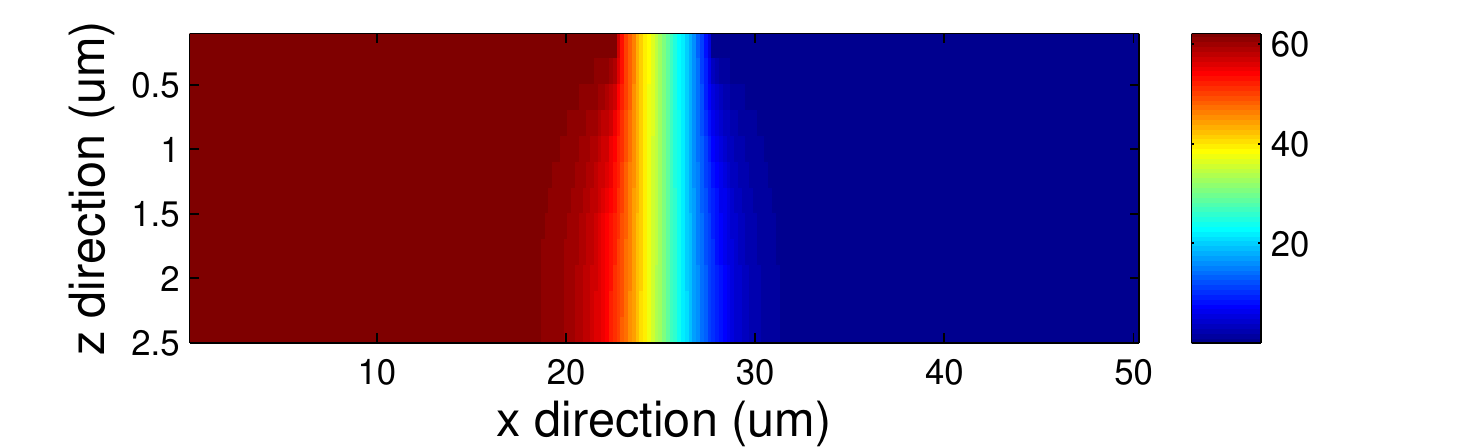}
\caption{Potential distribution in the $XOZ$ cross-section that passes through the middle of the PCA with $60V$ bias voltage.}
\label{fig:dc1}
\end{figure}

\begin{figure}[H]
\centering
\includegraphics[width=.8\textwidth]{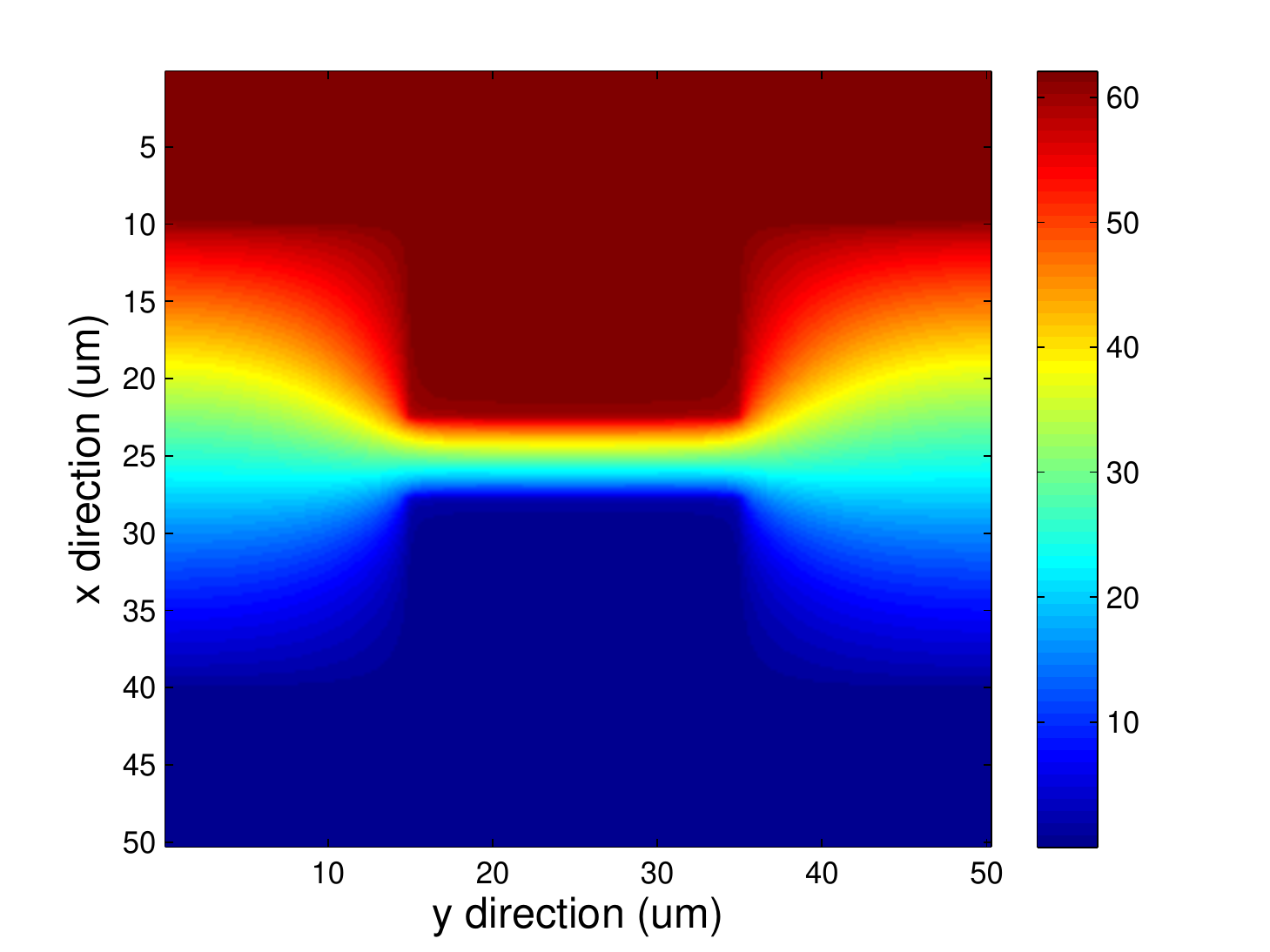}
\caption{Potential distribution on the top surface ($XOY$ cross-section) of the PCA with $60V$ bias voltage.}
\label{fig:dc2}
\end{figure}

The material and dimension parameters used for the simulation are summarized in Appendix B. For DC simulation, the step size of the voltage is $0.025V$.For FDTD simulation, the mesh size is $0.2\mu m$, and the time step is $0.33 fs$. All of simulations are carried out in 3D by Matlab. The steady-state solution of the potential distribution in the semiconductor with $60V$ bias voltage is shown in Fig.\ref{fig:dc1} and \ref{fig:dc2}. It can be seen that the dramatic variation happens inside the gap of the electrodes.

\begin{figure}[H]
\centering
\includegraphics[width=1\textwidth]{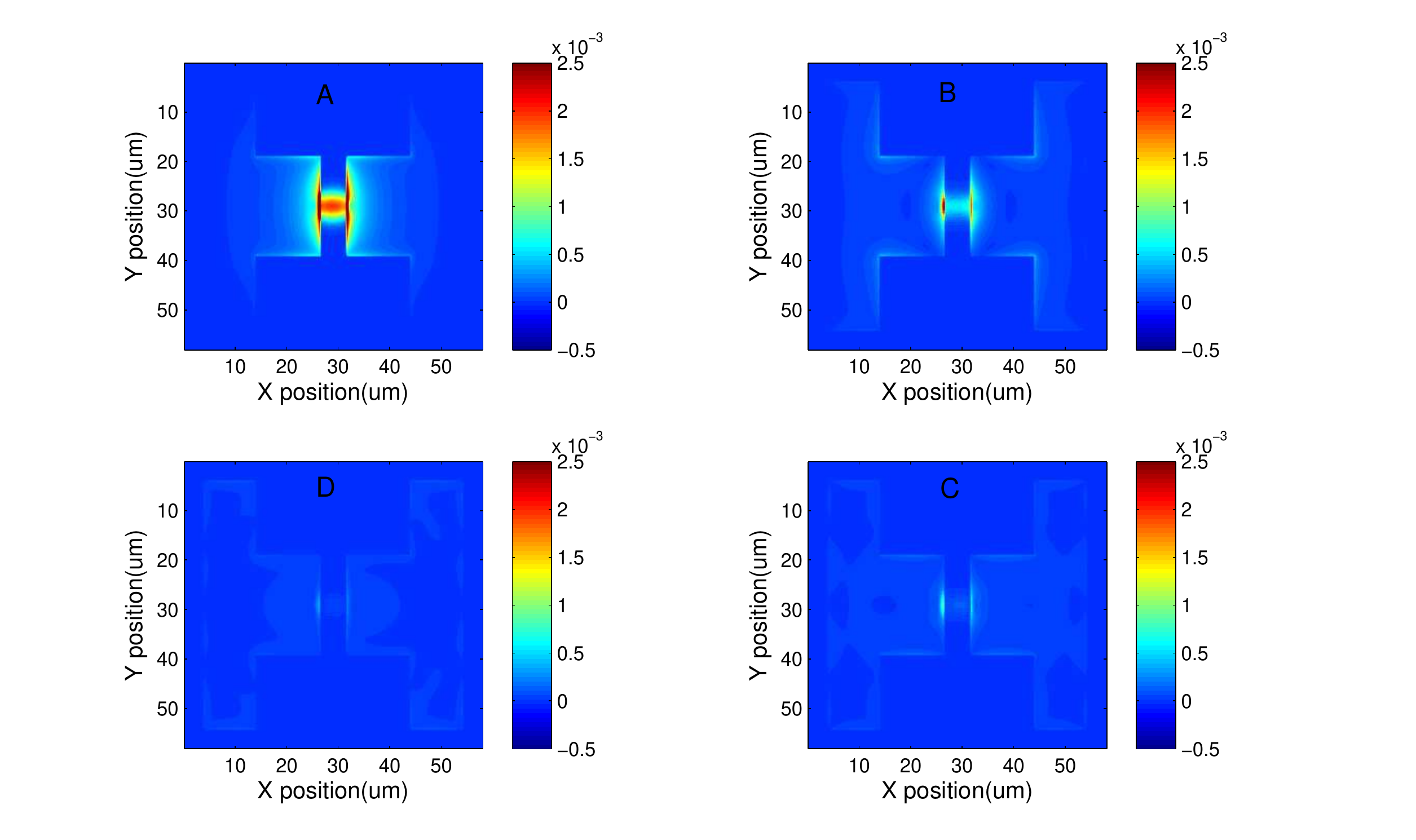}
\caption{Transient current density and electric field at the very top surface($XOY$ cross-section) of the PCA at (A) $0.5ps$ (the peak of laser pulse),(B)$0.9ps$,(C)$1.3ps$,and (D)$1.7ps$.}
\label{fig:ej}
\end{figure}

\begin{figure}[H]
\centering
\includegraphics[width=.8\textwidth]{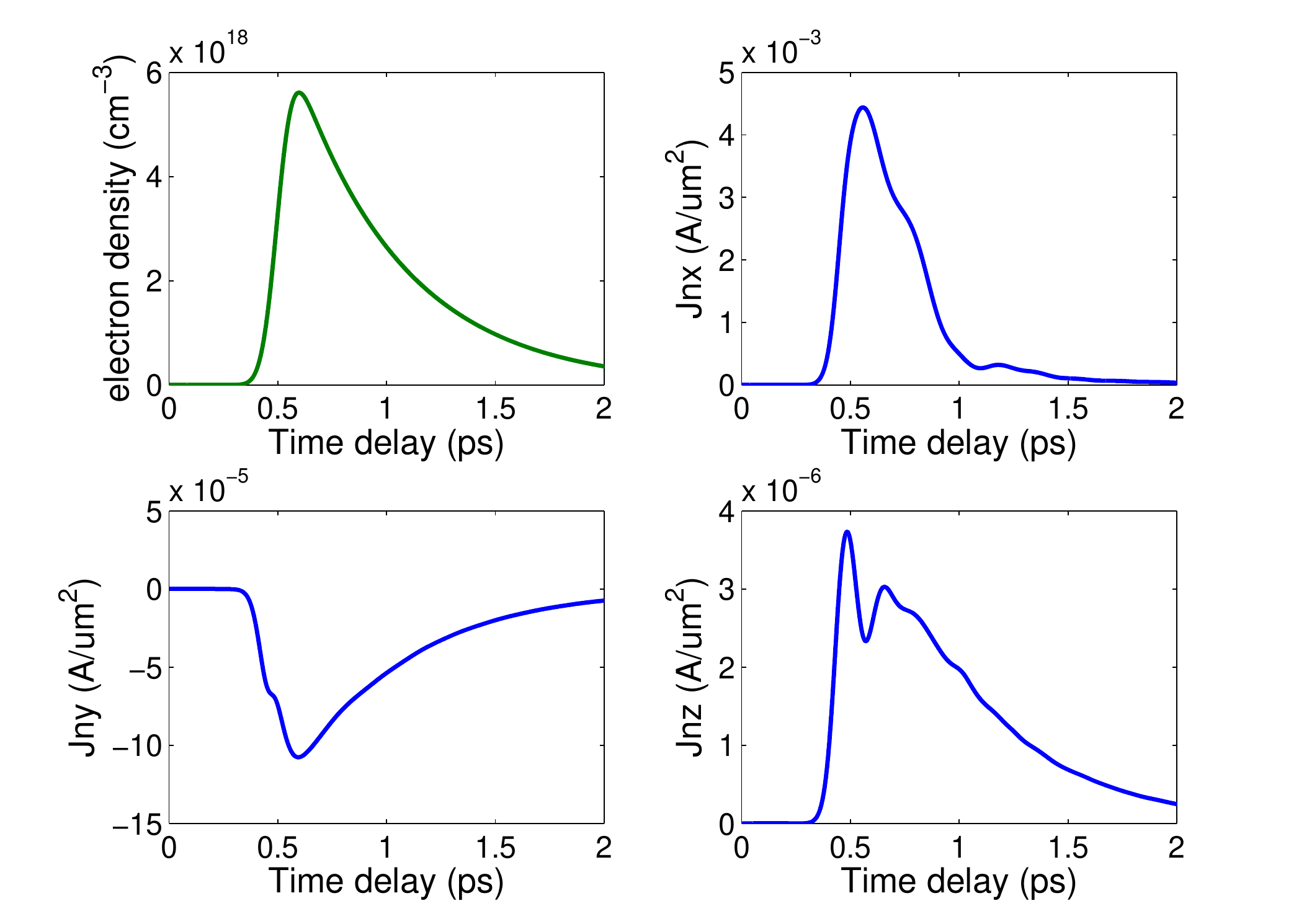}
\caption{Temporal behaviors of electron's density and current density at the center of the very top surface of the PCA.(a) Electron's density, (b) Current density in x-direction,(c) Current density in y-direction, and (d) Current density in z-direction.}
\label{fig:nj2ps}
\end{figure}

After the steady-state solution with biased voltage is obtained, the temporal behaviors are simulated for $2 ps$, within which the peak of the laser excitation is added at $0.5 ps$. The results are shown in Fig.4-6. Figure \ref{fig:ej} shows the transient values of current density and electric field at $0.7 ps$ at the very top surface ($XOY$ cross-section)of the PCA. Figure \ref{fig:nj2ps} shows the temporal behaviors of the photo-excited carrier density and the corresponding current density at the center (same as the location of the laser beam shown in Fig.\ref{fig:dimension}) of the very top surface of the PCA. It can be seen from Fig. \ref{fig:nj2ps} that the photo-excited current is dominant by the $x$-direction component, whose peak value is almost two orders of others.

\begin{figure}[H]
\centering
\includegraphics[width=.8\textwidth]{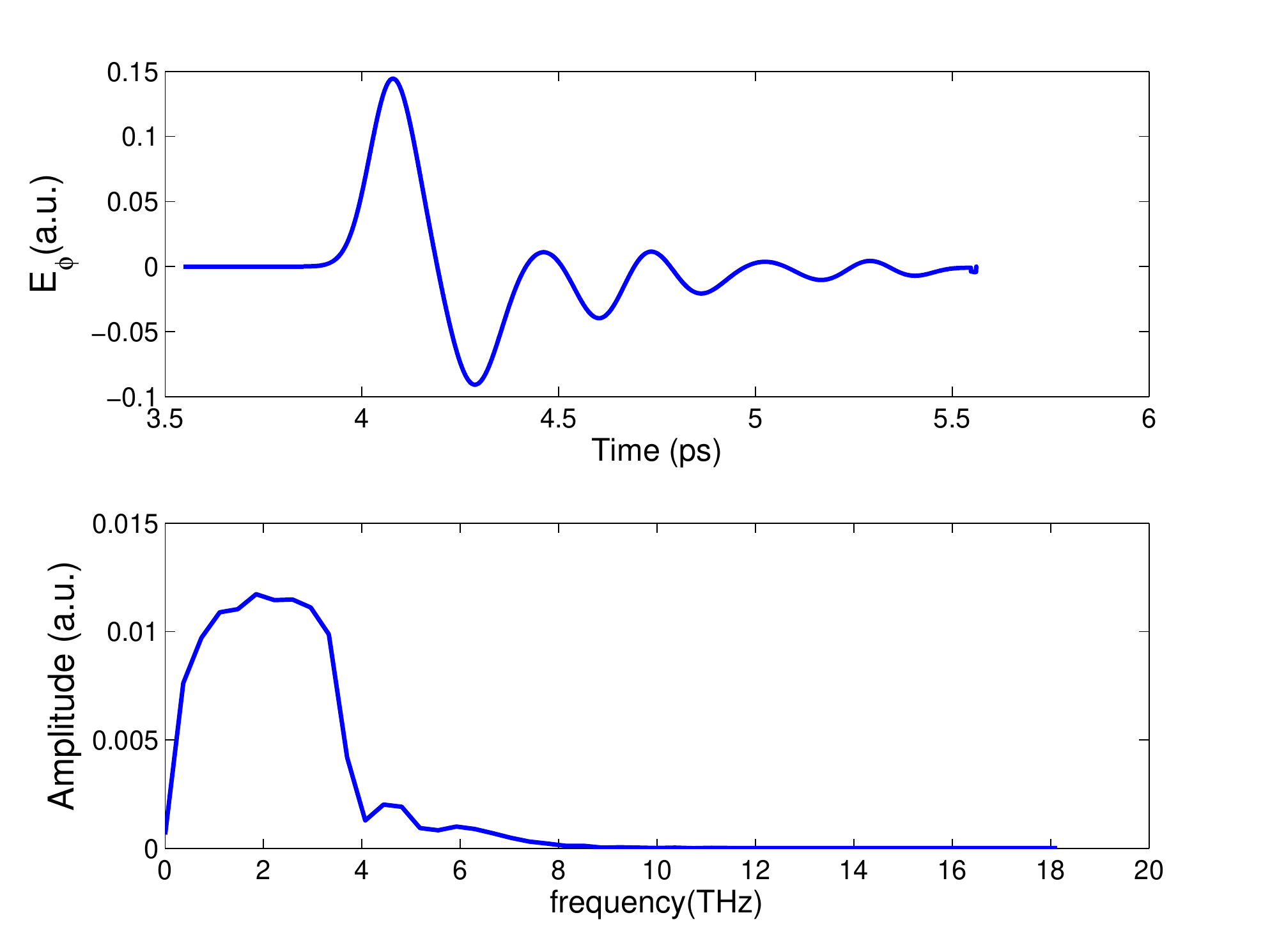}
\caption{Simulation result of the far-field radiation of the PCA.(upper)time-domain THz pulse, and (lower)the corresponding spectrum.}
\label{fig:far_field_E}
\end{figure}

\begin{figure}[H]
\centering
\includegraphics[width=.8\textwidth]{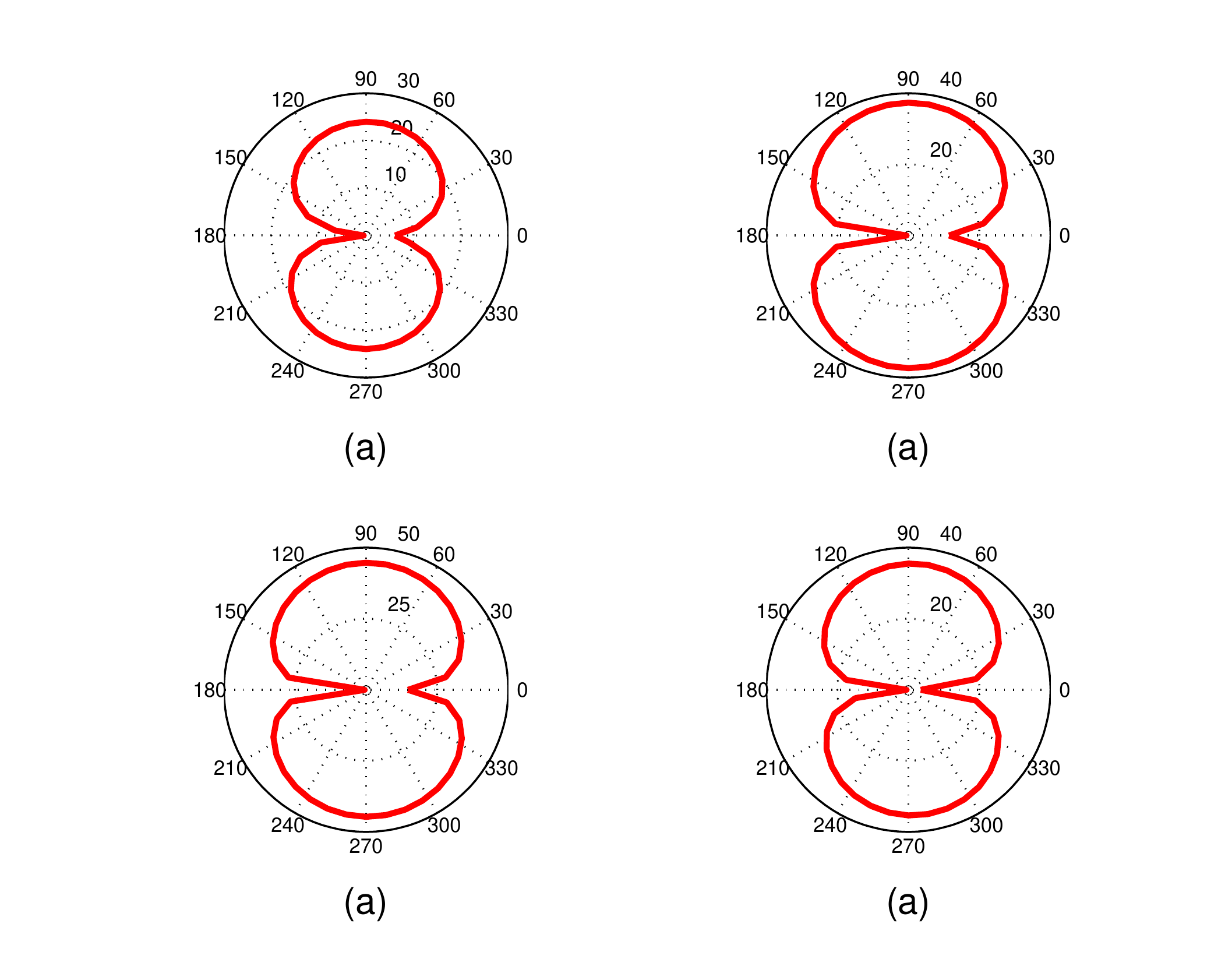}
\caption{Field pattern in far-field ($200 mm$ away from the center of PCA) at $XOZ$ plane.(a)$0.37 THz$,(b)$1.11 THz$,(c)$1.85 THz$, and (d) $2.59THz$.}
\label{fig:pattern_phi}
\end{figure}

\begin{figure}[H]
\centering
\includegraphics[width=.8\textwidth]{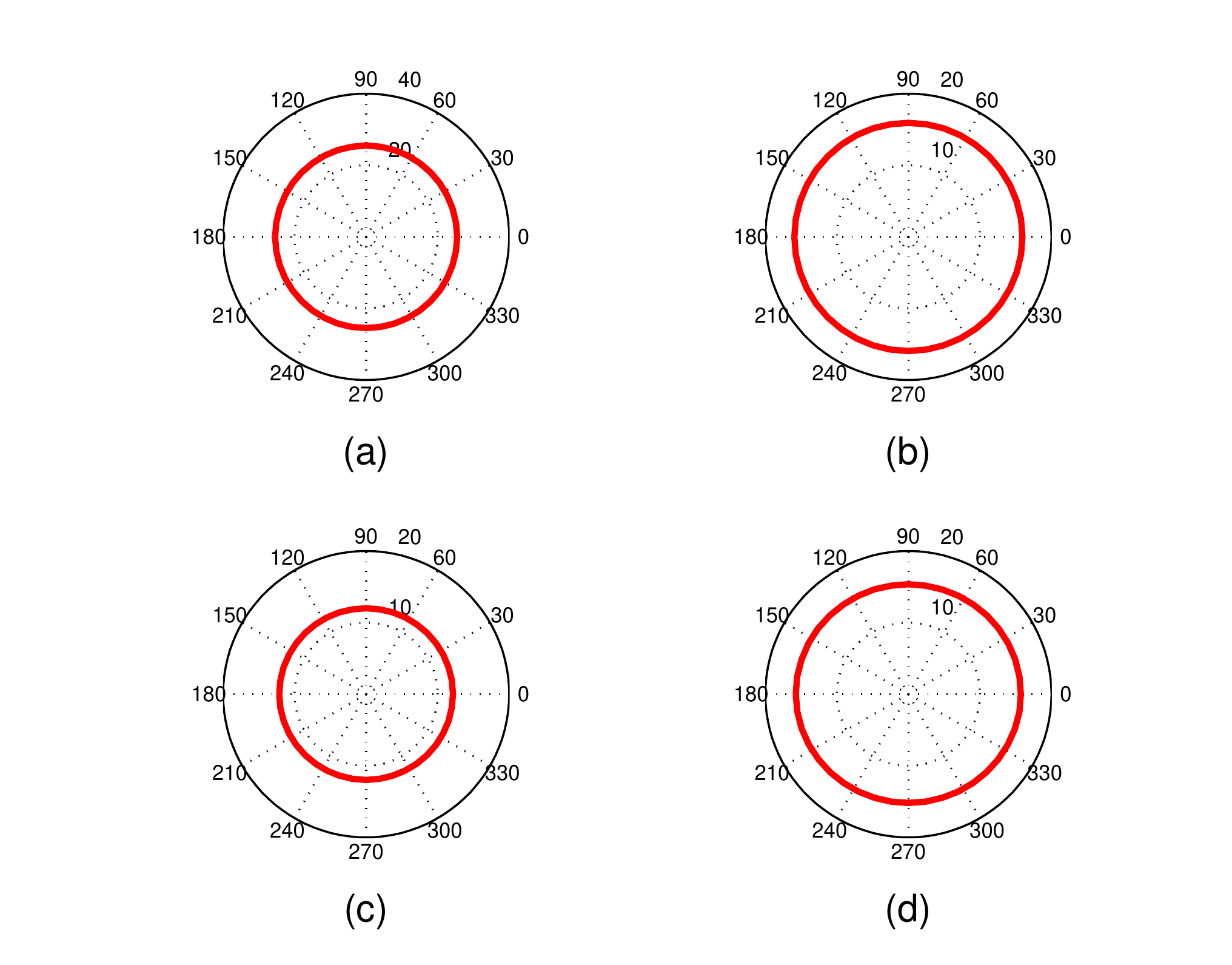}
\caption{Field pattern in far-field($200 mm$ away from the center of PCA) at $YOZ$ plane.(a)$0.37 THz$,(b)$1.11 THz$,(c)$1.85 THz$, and (d) $2.59THz$.}
\label{fig:pattern_theta}
\end{figure}

For far-field simulation, the far-field point is chosen right below the PCA with a distance of $200mm$, and the result is shown in Figure \ref{fig:far_field_E} (Since the far-field radiation is polarized along x-axis, only $E_\phi$ is shown). Beyond that, we also calculate the radiation pattern in far-field, and the results are shown in Fig.\ref{fig:pattern_phi}and\ref{fig:pattern_theta}. It is indicated that the radiation characters of the strip-line PCA is very close to that of a dipole.

\subsection{Parameter study result}

We implemented the parameter study of the PCA using the proposed simulation methods. The same coplanar strip-line PCA as that of Sec.4.1 is used in the simulation. The laser power is varied from $2.6mw$ to $60mw$ at the same bias voltage, and the corresponding far-field radiations are simulated. The result is shown in Fig.\ref{fig:far_field_E}. Along with the increase of the laser power, the radiated THz field will increase monotonously but show saturation effect at high laser power. This phenomena has been observed in experiment and can be explained by the scaling rule \citep{darrow1992}. Other parameter studies will be reported elsewhere in future.

\begin{figure}[H]
\centering
\includegraphics[width=.8\textwidth]{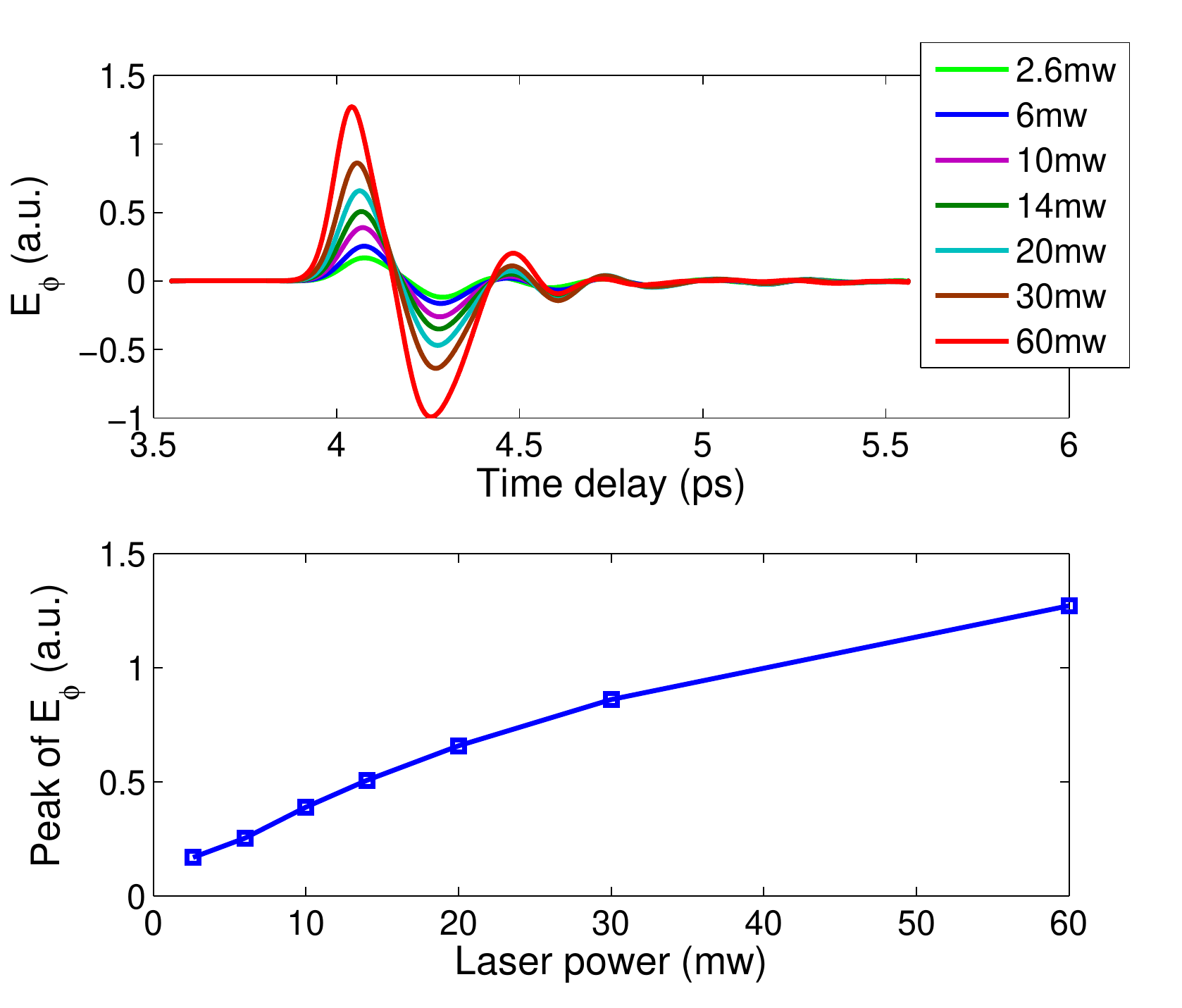}
\caption{Dependence of THz field on the laser power.(upper)Time-domain THz pulse in far-field,(lower)Relationship between the peak of THz pulses and laser powers. Squres represent simulated data and the solid curve is a guide of the eye.}
\label{fig:parameter_Evspower}
\end{figure}

\section{Conclusion}

A full-wave simulation method is developed to simulate the response of the THz PCA in both the near-field and far-field. The validity of this method has been verified by comparing with commercial softwares. The advantage of this method is that it couples multi-physics together so that is capable of characterizing the PCA comprehensively. Owing to this method, the performance of a PCA under various conditions can be predicated and evaluated. Furthermore, the novel design of PCA that promising better performance (such as higher THz radiation power and optics-to-THz efficiency) can be implemented based on this method.

\section*{Acknowledgements}
This work can not be done without the contribution of the following colleagues. They are Mingguang Tuo, Min Liang and Hao Xin. They will be listed as co-authors when we consider to publish this work in a journal.

\appendix
\section{Discretization of time-domain equations }
In $x$-direction, the partial differential equations can be discretized as follows

\begin{align}
H_{x}\arrowvert _{i,j+1/2,k+1/2}^{t+1/2}= & \notag H_{x}\arrowvert _{i,j+1/2,k+1/2}^{t-1/2} + \frac{\bigtriangleup t}{\mu \cdot \bigtriangleup x}\cdot \\
& \left[E_{z}\arrowvert _{i,j+1,k+1/2}^{t} - E_{z}\arrowvert _{i,j,k+1/2}^{t} \right.\nonumber \\
&\qquad \left.{} + E_{y}\arrowvert _{i,j+1/2,k+1}^{t} - E_{y}\arrowvert _{i,j+1/2,k}^{t} \right]
\end{align}

\begin{align}
E_{x}\arrowvert _{i+1/2,j,k}^{t+1}= & \notag E_{x}\arrowvert _{i+1/2,j,k}^{t} + \frac{\bigtriangleup t}{\varepsilon \cdot \bigtriangleup x}\cdot \\
& \left[H_{z}\arrowvert _{i+1/2,j+1,k}^{t+1/2} - H_{z}\arrowvert _{i+1/2,j,k}^{t+1/2} \right.\nonumber \\
& + H_{y}\arrowvert _{i+1/2,j,k+1}^{t+1/2} - H_{y}\arrowvert _{i+1/2,j,k}^{t+1/2} \\
&\notag \qquad \left.{} -J_{nx}\arrowvert _{i+1/2,j,k}^{t+1/2}-J_{px}\arrowvert _{i+1/2,j,k}^{t+1/2} \right]
\end{align}

\begin{align}
n\arrowvert _{i,j,k}^{t+1/2}= & \notag n\arrowvert _{i,j,k}^{t-1/2} + \frac{\bigtriangleup t}{q \cdot \bigtriangleup x}\cdot \\
& \left[J_{nx}\arrowvert _{i+1/2,j,k}^{t} - J_{nx}\arrowvert _{i-1/2,j,k}^{t} \right.\nonumber \\
&\notag + J_{ny}\arrowvert _{i,j+1/2,k}^{t} - J_{ny}\arrowvert _{i,j-1/2,k}^{t} \\
&\notag + J_{nz}\arrowvert _{i,j,k+1/2}^{t} - J_{nz}\arrowvert _{i,j,k-1/2}^{t} \\
& \qquad \left.{} +q\cdot (G\arrowvert _{i,j,k}^{t}-R\arrowvert _{i,j,k}^{t}) \right]
\end{align}

\begin{align}
p\arrowvert _{i,j,k}^{t+1/2}= & \notag p\arrowvert _{i,j,k}^{t-1/2} - \frac{\bigtriangleup t}{q \cdot \bigtriangleup x}\cdot \\
& \left[J_{px}\arrowvert _{i+1/2,j,k}^{t} - J_{px}\arrowvert _{i-1/2,j,k}^{t} \right.\nonumber \\
&\notag + J_{py}\arrowvert _{i,j+1/2,k}^{t} - J_{py}\arrowvert _{i,j-1/2,k}^{t} \\
&\notag + J_{pz}\arrowvert _{i,j,k+1/2}^{t} - J_{pz}\arrowvert _{i,j,k-1/2}^{t} \\
& \qquad \left.{} -q\cdot (G\arrowvert _{i,j,k}^{t}-R\arrowvert _{i,j,k}^{t}) \right]
\end{align}

\begin{align}
J_{nx}\arrowvert _{i+1/2,j,k}^{t}= & \notag q \mu_{n}\cdot n\arrowvert _{i+1/2,j,k}^{t} \cdot \left[E_{DCx}\arrowvert _{i+1/2,j,k}+E_{x}\arrowvert _{i+1/2,j,k}^{t}\right] \\
& + qD_{n} \cdot \frac{n\arrowvert _{i+1,j,k}^{t} - n\arrowvert _{i,j,k}^{t} }{\bigtriangleup x} 
\end{align}

\begin{align}
J_{px}\arrowvert _{i+1/2,j,k}^{t}= & \notag q \mu_{p}\cdot p\arrowvert _{i+1/2,j,k}^{t} \cdot \left[E_{DCx}\arrowvert _{i+1/2,j,k}+E_{x}\arrowvert _{i+1/2,j,k}^{t}\right] \\
& - qD_{p} \cdot \frac{p\arrowvert _{i+1,j,k}^{t} - p\arrowvert _{i,j,k}^{t} }{\bigtriangleup x} 
\end{align}
\noindent where $(i,j,k)$ represents space location, $t$ represents current time-loop, and $\bigtriangleup x$ and $\bigtriangleup t$ represent space step and time step, respectively. The discretization forms of the components in other directions can be deduced accordingly.

\section{Summary of the input data in the simulation}
The dimension of the PCA is shown in the figure below, and other parameters used in the simulation are summarized in Table 1. For the sake of simplicity, the top surface of the semiconductor is also 50$\mu m\times$ 50$\mu m$, and the thickness is 2.2 $\mu m$.

\begin{figure}[H]
\centering
\includegraphics[width=.8\textwidth]{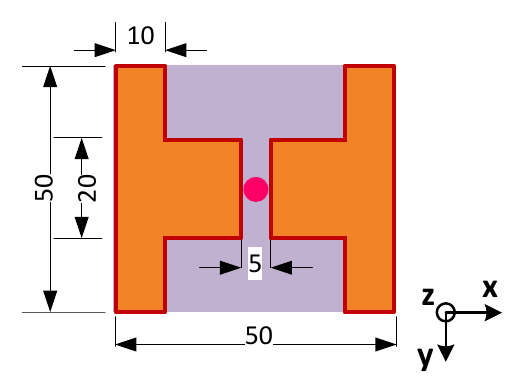}
\caption{Dimension of the PCA, all dimensions are in micrometer. The solid circle indicates the location of the laser beam.}
\label{fig:dimension}
\end{figure}

\begin{table}[H]
\caption{Parameters used in the simulation}
\centering
    \begin{tabular}{ll}
    \hline
    \textbf{Parameters}                        & \textbf{Values}                   \\ \hline
    Material                          & LT-GaAs                  \\
    Carrier lifetime ($ps$)             & electron: 0.1, hole: 0.4 \\
    Mobility ($cm^{2}/V\cdot s$) & electron: 200, hole: 30  \\
    Permittivity                      & 12.9                     \\
    Intrinsic concentration ($cm^{-3}$) & 2.1E6                      \\
    Absorption coefficient ($cm^{-1}$) & 1E4                     \\
    Laser wavelength ($nm$)             & 800                      \\
    Beam waist ($\mu m$)       & 2.5                  \\
    Pulse duration ($fs$)               & 80                       \\
    Intensity ($W/cm^{2}$)           & 1E9                      \\
    DC voltage ($V$)           & 60                      \\ \hline
    \end{tabular}
\end{table}

 

\bibliographystyle{plain}
\bibliography{references}
\end{document}